\documentclass[letterpaper, 10 pt, conference]{ieeeconf}  

\IEEEoverridecommandlockouts                              

\usepackage{graphics} 
\usepackage{epsfig} 
\usepackage{mathptmx} 
\usepackage{times} 
\usepackage{amsmath} 
\usepackage{amssymb}  
\usepackage{pdfpages}
\usepackage{hyperref}
\usepackage{float}

\title{\LARGE \bf
Sparse-Denoising Methods for Extracting Desaturation Transients in Cerebral Oxygenation Signals of Preterm Infants*
}

\author{Minoo Ashoori$^{1,2}$, Eugene M. Dempsey$^{1,3}$, Fiona B. McDonald$^{1,2}$, and John M. O’ Toole$^{1,3}$ \textit{Member, IEEE}
\thanks{*This work was supported by Science Foundation Ireland (SFI 18/SIRG/5483) and (SFI 15/SIRG/3580).}
\thanks{1. Irish Centre for Maternal and Child Health Research (INFANT), University College Cork, Ireland. (email: {\tt\small 120224294@umail.ucc.ie})}%
\thanks{2. Department of Physiology, University College Cork, Ireland}%
\thanks{3. Department of Paediatrics and Child Health, University College Cork, Ireland}%
}

\begin{document}

\maketitle
\thispagestyle{empty}
\pagestyle{empty}

\begin{abstract}

Preterm infants are at high risk of developing brain injury in the first days of life as a consequence of poor cerebral oxygen delivery. Near-infrared spectroscopy (NIRS) is an established technology developed to monitor regional tissue oxygenation. Detailed waveform analysis of the cerebral NIRS signal could improve the clinical utility of this method in accurately predicting brain injury. Frequent transient cerebral oxygen desaturations are commonly observed in extremely preterm infants, yet their clinical significance remains unclear. The aim of this study was to examine and compare the performance of two distinct approaches in isolating and extracting transient deflections within NIRS signals. We optimized three different simultaneous low-pass filtering and total variation denoising (LPF--TVD) methods and compared their performance with a recently proposed method that uses singular-spectrum analysis and the discrete cosine transform (SSA--DCT). Parameters for the LPF--TVD methods were optimized over a grid search using synthetic NIRS-like signals. The SSA--DCT method was modified with a post-processing procedure to increase sparsity in the extracted components. Our analysis, using a synthetic NIRS-like dataset, showed that a LPF--TVD method outperformed the modified SSA--DCT method: median mean-squared error of $0.97$ ($95\%$ CI: $0.86$ to $1.07$) was lower for the LPF--TVD method compared to the modified SSA--DCT method of $1.48$ ($95\%$ CI: $1.33$ to $1.63$), $P < 0.001$. The dual low-pass filter and total variation denoising methods are considerably more computational efficient, by 3 to 4 orders of magnitude, than the SSA--DCT method. More research is needed to examine the efficacy of these methods in extracting oxygen desaturation in real NIRS signals.
\newline

\indent \textit{Clinical relevance}— Early and precise identification of abnormal brain oxygenation in premature infants would enable clinicians to employ therapeutic strategies that seek to prevent brain injury and long-term morbidity in this vulnerable population. 
\end{abstract}

\section{INTRODUCTION}

Near-infrared spectroscopy (NIRS) is a non-invasive optical technology used for continuous monitoring of cerebral regional oxygen saturation \cite{c1}. Despite the widespread use of NIRS in the neonatal intensive care unit, ranges remain poorly described for extremely preterm infants. In an ongoing clinical trial \cite{c2} to evaluate the use of NIRS-guided treatment of extremely preterm infants, short duration desaturations are commonly observed in the NIRS signal. Isolating these transient deflection within the signals without removing other components of the signal is an important aspect of processing these signals for post-hoc analysis \cite{c3}.

A recent study developed a novel method to isolate transient waveforms in NIRS signals and showed that these transient components of the signal are not predictive of brain injury \cite{c4}, however there has been no detailed investigation of the most suitable method for extracting these short duration desaturations. Effective and efficient decomposition of these short duration desaturations would provide an excellent opportunity to test their clinical utility. The aim of this study is to identify the best method to isolate and extract transient deflections within synthetic NIRS signals. We optimize two distinct methods independently; singular-spectrum analysis and the discrete cosine transform (SSA-–DCT) method (SSA--DCT1 and SSA--DCT2) \cite{c4} and low-pass filtering total variation denoising (LPF--TVD) method (LPF--TVD1, LPF--CS1, LPF--CSD2) \cite{c5, c6}. We then compare their performance in isolating and extracting transient deflections within NIRS-like signals.
 
 \begin{figure*}[thbp]
      \centering
      \includegraphics[width=1\linewidth,height=0.5\textheight, scale=0.1]{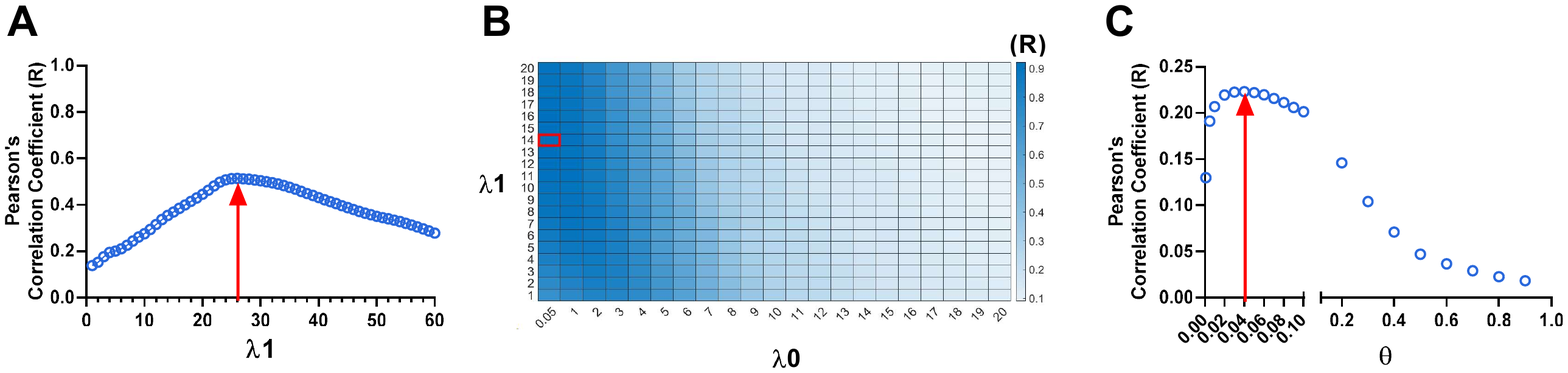}
      \vspace*{-85mm} 
      \caption{Grid search for the parameters of the LPF--TVD methods; A. Correlation coefficient (R) for values of $\lambda_{1}$ in LPF--TVD1; B. Correlation coefficient for values of $\lambda_{0}$ and $\lambda_{1}$ in LPF--CSD1; C. Correlation coefficient for values of $\theta$ in LPF--CSD2; Arrow denotes maximum value of R; LPF--TVD1 -- low-pass filtering total variation denoising; LPF--CSD1 -- low-pass filtering compound sparse denoising; LPF--CSD2 -- the improved version of LPF--CSD1 }
      \vspace*{-5mm}
      \label{figurelabel}
   \end{figure*}
   
\vspace*{-1mm}   
\section{METHODS}

We present a simple post-processing method to increase sparsity of the SSA--DCT decomposition method \cite{c4} and compared this modified algorithm with multiple algorithms of LPF--TVD. To validate and compare these methods, we use a set of synthetic rcSO$_{2}$ (regional cerebral oxygenation) signals with various numbers of transients \cite{c4}. 

\subsection{Singular-Spectrum Analysis for Extracting Transients}
\label{sec:sing-spectr-analys}

\subsubsection{Singular-Spectrum Analysis}

The SSA method is presented as a data-driven filter-bank operation. 
For signal $\mathbf{x} = x[n]$ for $n = 0, 1, \ldots, N-1$, we form the $k$-th lagged vector with
embedding dimension $M$ as $\mathbf{x}_k = \{x[k+M-1-m]\}^T$ for $m=1,2,\ldots,M$.
Combining these lagged vectors, for $k=1,\ldots,K$ with $K=N-M+1$, we form the trajectory
$M\times K$ matrix as $\mathbf{X} = (\mathbf{x}_1,\mathbf{x}_2,\ldots,\mathbf{x}_K)$ \cite{c7}.
  The correlation matrix of the trajectory matrix is decomposed into an orthogonal basis
  using eigenvalue decomposition:
\begin{equation*}
  \mathbf{R} = \mathbf{X}\mathbf{X}^T = \mathbf{U}\mathbf{\lambda}\mathbf{U} 
\end{equation*}

Matrix $\mathbf{U} =(\mathbf{u}_1,\ldots,\mathbf{u}_M)^T$ contains the basis vectors
$\mathbf{u}_m$, known as eigenvectors; matrix $\mathbf{\lambda}$ =
diag$(\lambda_1,\ldots,\lambda_M)$ contains the weights $\lambda_m$, known as eigenvalues.  
Each component $\mathbf{x}_m$ is generated through a finite-impulse response (FIR) filter
with the eigenvector $\mathbf{u}_m$ as the impulse response as
\begin{equation*}
  \mathbf{x}_m = \mathcal{F}^{-1}\{ \mathcal{F}\{\mathbf{x}\} |\mathcal{F}\{ \mathbf{u}_m \}|^2 \}.  
\end{equation*}

Here, $\mathcal{F}$ represents the discrete-time Fourier transform, and $\mathcal{F}^{-1}$ indicates the
inverse transform of $\mathcal{F}$.  Thus, the SSA method decomposes signal $\mathbf{x}$
into $M$ components $\mathbf{x}_m$, where $\mathbf{x}=\sum_{m=1}^{M}\mathbf{x}_m$.  

Rejecting components $\mathbf{x}_m$ associated with noise can increase the signal-to-noise
ratio of $\hat{\mathbf{x}}$ \cite{c7, c8}, where
$\hat{\mathbf{x}}=\sum_{m\in D} \mathbf{x}_m$ and $D$ is a subset of $\{1,2\ldots,M\}$. 
There are different methods to select this subset $D$, usually based on the values of the
associated eigenvalues. 
Here, we used a method proposed by Vautard \emph{et al.} \cite{c8} based on a previous
comparison of different methods for the same class of signals \cite{c4}. 
The method compares the possible noise components $\sum_{m \not\in D} \mathbf{x}_m$ in the
autocorrelation domain to upper limits of a white Gaussian noise process. These
upper-limits are estimated from a $95\%$ confidence interval generated from $100$ iterations
of white Gaussian noise.  

\subsubsection{Extracting transients}

The signal $\mathbf{x}$ is transformed using the discrete cosine transform (DCT). 
This equates to a $90^{\circ}$ rotation in the time--frequency plane. 
Therefore impulses, or impulse-like transients, are transformed to sinusoidal-type
components. 
Thus, applying the SSA to the real-valued DCT signal enables the algorithm to extract
oscillating components, which when transformed back to the time-domain through an inverse
DCT, represents transient components. 
This decomposition was refined through an iterative procedure. 
Full details can be found in \cite{c4}.

\subsubsection{Post-processing}

Although initial results were promising for the SSA--DCT method \cite{c4}, we
found the method did not perform well when applied to long duration cerebral oxygenation
signals. 
In particular, we found the extracted transient components included a wandering baseline
that distorted the rcSO$_2$ signal. To rectify this, we applied a simple post-processing
procedure that generates a mask, bounded within $[0,1]$, to limit the wandering baseline
by multiplying the mask with the extract transient component. We refer to this modified method as SSA--DCT2 and the original method as SSA--DCT1.

The mask was generated as follows. 
First, a threshold is applied to the extracted transient component $\mathbf{x}_t$, so that
the mask $m[n]=1$ when $x_t[n]<-T$; otherwise $m[n]=0$. This ensures that transient
components must be at $>T$ (\% rcSO$_2$) in amplitude and always negative. 
Second, the sharp edges of the binary mask were expanded using one-half of a
10 minute Blackman-Harris window to enable a smooth transition from $0$ to $1$. Lastly,
the mask was bounded to $[0,1]$, as overlapping transitions may have increased $m[n]>1$.  
From initial testing, $T$ was set to $8$ (\% rcSO$_2$), although
this could be changed in future iterations based on physiological definitions of cerebral
desaturations. Matlab code for both SSA–-DCT methods are freely available at \url{https://github.com/otoolej/transient_decomp_ssa} (version 0.2.0 ).

\subsection{Total Variation Denoising Methods}
\label{sec:total-vari-denos}

Total variation denosing (TVD) is a sparsity-based denoising method that estimates a signal component with a sparse derivative. TVD methods can be extended to enforce sparsity on the signal itself, in addition to the derivative. Here we examine methods that combine low-pass filtering with TVD (LPF–-TVD) to allow for the decomposition of the signal into a low-pass component in addition to a sparse component. This process differs to low-pass filtering the signal first, which can dampen transients \cite{c5}. Given signal $\mathbf{y}$, we find
$\mathbf{x}$ by minimizing the following cost function
\begin{equation}
  \label{eq:2}
  \arg \underset{\mathbf{x}}{\min} \left\{\frac{1}{2}||\mathbf{H}(\mathbf{y} -
    \mathbf{x})||_2^2
    + \lambda_0 ||\mathbf{x}||_1 + \lambda_1 ||\mathbf{D}\mathbf{x}||_1 \right\}
\end{equation}
where $||x||_p = (\sum_{n}|x[n]|^p)^{1/p}$ represents the $l_p$ norm for $\mathbf{x}$
and $\mathbf{D}$ represents the finite-difference matrix, such that
$||\mathbf{D}\mathbf{x}||_1 = \sum_n |x[n+1] -x[n]|$.  Matrix $\mathbf{H}$ is a high-pass
filter.
Regularization parameters $\lambda_0$ and $\lambda_1$ are application dependent.

For the first method, which we refer to as LPF--TVD1, $\lambda_0=0$. 
Thus the signal $\hat{\mathbf{x}}$ from \eqref{eq:2} may not be sparse but will have a
sparse derivative.
This minimization problem in \eqref{eq:2} is solved using a majorization--minimization
convex optimization algorithm. 
The second method, known as the low pass filtering compound sparse denoising (LPF--CSD1) method, estimates
$\hat{\mathbf{x}}$ from \eqref{eq:2} using another convex optimization algorithm, the
alternating direction method of multipliers.
Further details on both algorithms can be found in \cite{c5}.

The third method is an improved version of the LPF--CSD1 method
\cite{c6}. 
This method, which we call LPF--CSD2, uses the optimization problem in
\eqref{eq:2} but replaces the $l_1$ norm with non-convex penalty functions $\phi_0$ and $\phi_1$
\begin{multline}
  \label{eq:5}
  \arg \underset{\mathbf{x}}{\min} \left\{\frac{1}{2}
    ||\mathbf{H}(\mathbf{y} - \mathbf{x})||_2^2 + \lambda_0 \sum_{n=0}^{N-1}\phi_0(x[n]) \right.  \\
      \left. + \lambda_1 \sum_{n=0}^{N-1}\phi_1(x[n+1]-x[n]) \right\}
\end{multline}

The use of these penalty functions allows for a better estimate of transient components as they are less biased towards zero \cite{c6}. 
We used the following function, with parameter $r$, for $\phi_0$ and $\phi_1$
\begin{equation}
  \label{eq:4}
  \phi(u; r) = \frac{2}{r\sqrt{3}}
  \left[\tan^{-1}\left(\frac{1+2r\sqrt{u^2+\epsilon}}{\sqrt{3}}\right) - \frac{\pi}{6} \right]
\end{equation}
with $\epsilon$ set to $10^{-10}$ to avoid singularities at $u=0$ in the derivative of $\phi$.  


\begin{figure}[H]
      \centering
       \vspace*{-5mm} 
      \includegraphics[width=1\linewidth,height=0.4\textheight,scale=0.6]{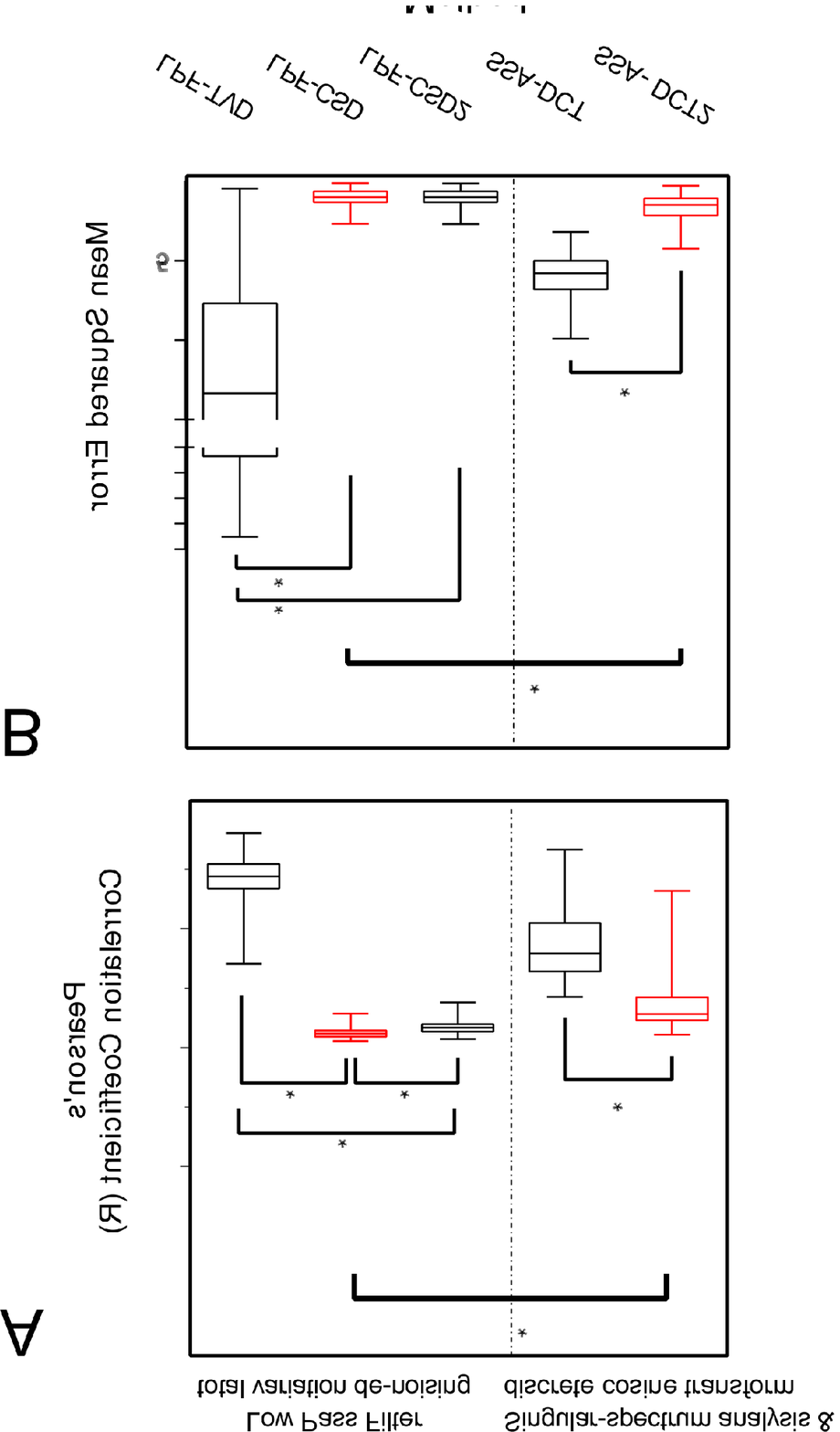}
      \vspace*{-10mm}
      \caption{Comparison between the LPF--TVD methods and between the SSA--DCT methods, and comparison between the best performing algorithm in each group when assessed on 100 synthetic NIRS-like signals; * indicates significant difference; LPF--TVD1 -- low-pass filtering total variation denoising; LPF--CSD1 -- low-pass filtering compound sparse denoising; LPF--CSD2 -- the improved version of LPF--CSD1; SSA–-DCT1 -- singular-spectrum analysis and the discrete cosine transform; SSA–-DCT2 -- SSA--DCT1 with post-processing}
      \label{figurelabel}
   \end{figure}

\subsection{Performance Comparison}
\label{sec:comparing-methods}

To compare methods, we used $100$ synthetic rcSO$_2$-like signals of length 43,200 sample points, corresponding to $72$ hours with sampling frequency of $1/6$ Hz, typical of our NIRS recordings over the first days of life. The synthetic signals are a linear combination of transient components and nonstationary colored noise; full details can be found in \cite{c4}. The number and amplitude of the transient components were adjusted such that they were representative of NIRS signals of extremely pre-term infants ($<28$ weeks gestational age).

For the three LPF--TVD methods, we optimized the parameters for each method using a grid search.  
Each method has a different set of parameters: LPF--TVD1 has $\lambda_1$; LPF--CSD1 has
$\lambda_0$ and $\lambda_1$; LPF--CSD2 has $\lambda_{0}$, $\lambda_{1}$, $r_0$ and $r_1$, as parameters of the penalty functions $\phi_0$ and $\phi_1$ in \eqref{eq:5} and \eqref{eq:4}. Parameters $\lambda_0$ and $\lambda_1$ in LPF--CSD2 were estimated using the procedure proposed in [6]. The correlation coefficient between the estimated and pre-defined transient component was used as the optimization metric. A different set of $50$ synthetic rcSO$_2$ of length 12-hours were used in the grid search. Based on initial testing on a subset of the data, we used a 1st order high-pass filter,
$\mathbf{H}$ in \eqref{eq:2} and \eqref{eq:5}, with a normalized cut-off frequency $fc = 0.01$. 
The 5 methods were compared with each other using Kruskal-Wallis test utilizing the selected parameters from the grid search on the data set of $100$ rcSO$_2$-like signals. Dunn’s post-hoc test was performed to compare the three LPF--TVD methods with each other, SSA--DCT1 versus SSA--DCT2 as well as the best performing SSA-–DCT method versus the best performing  LPF-–TVD method.

Methods were compared using two metrics---the mean-squared error and correlation
coefficient between the pre-defined and estimated transient components---to test the performance at estimating the transient components of the synthetic signals. Statistical significance was determined as $P<0.05$.

\begin{figure*}[thbp]
\centering
\includegraphics[width=0.9\linewidth,height=0.28\textheight,scale=1]{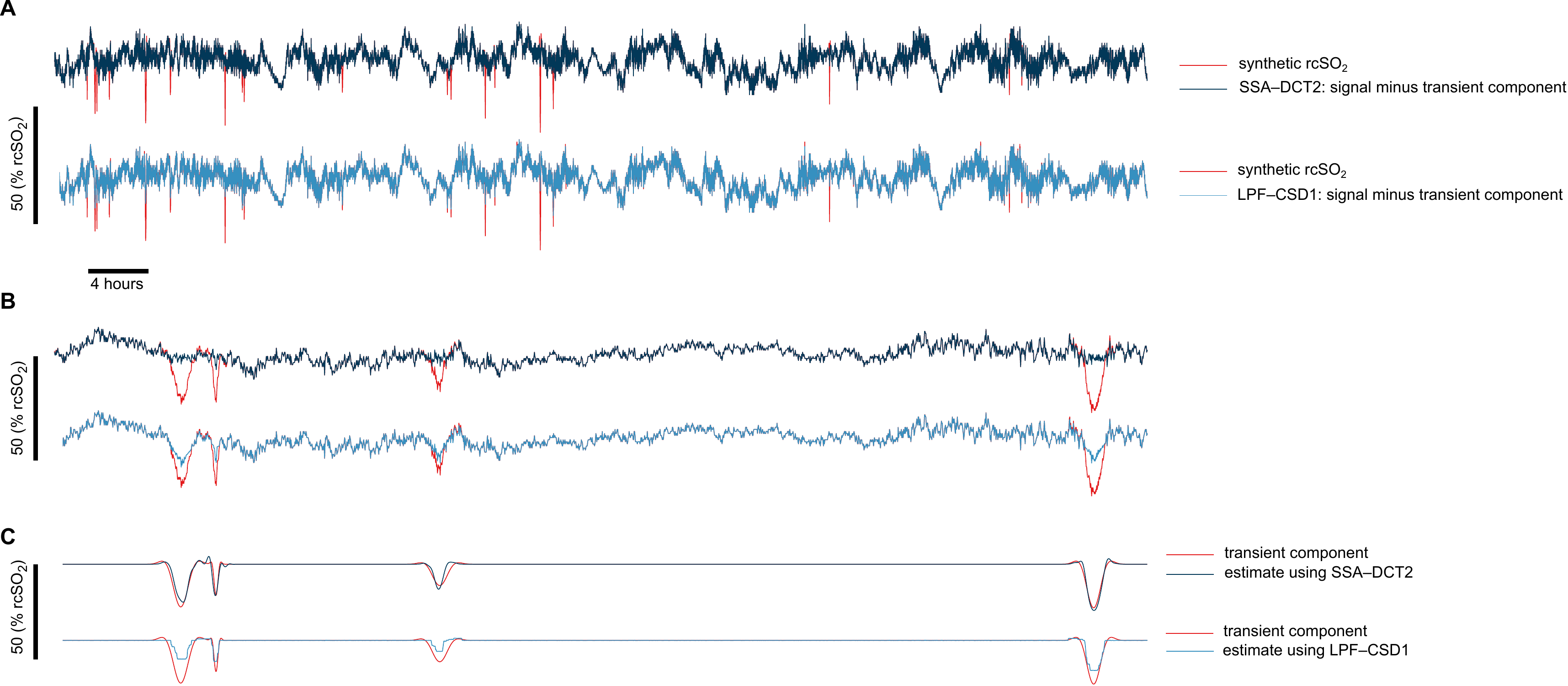}  
 \vspace*{-3mm} 
\caption{A. SSA--DCT2 and LPF--CSD1 decomposition methods on a NIRS-like synthetic signal; B. Zoomed-in area from A, indicated by the ‘4 hours’ line; c. Zoomed-in predefined transients and extracted ones; rcSO2 -- regional cerebral oxygen saturation; SSA--DCT2 -- singular-spectrum analysis and the discrete cosine transform with post-processing; LPF--CSD1 -- low-pass filtering compound sparse denoising }
\vspace*{-5mm} 
\label{fig:ssa_eg}
\end{figure*}

  \vspace*{-3mm} 
\section{RESULTS}

Grid search for the LPF--TVD parameters found the following. For the LPF--TVD1 method $\lambda_{1} = 26$ (Fig. 1A); for the LPF--CSD1 method, the best combination of $\lambda_{0}$ and $\lambda_{1}$ was $\lambda_{0} = 0.5$ and $\lambda_{1} = 14$ (Fig. 1B); for the LPF-–CSD2 method, we use a standard deviation (SD) estimate of the noise component as $4.69$ and an optimized $\theta$ value of $0.04$, which results in  $\lambda_{0} = 0.54$, $\lambda_{1} = 21.32$, $r_0 = 0.001$ and $r_1 = 0.17$ (Fig. 1C) \eqref{eq:4}.

A comparison between the three LPF--TVD methods showed that the LPF--CSD1 method outperformed the other two methods- Fig. 2. LPF--CSD1 had statistically significantly higher correlation coefficient compared to LPF--CSD2: median correlation for LPF--CSD1 was $0.94$ ($95\%$ CI: $0.93$ to $0.94$), compared to median correlation of $0.92$ ($95\%$ CI: $0.91$ to $0.93$) for LPF--CSD2, $P = 0.001$. Moreover, LPF-TVD1 method had the lowest performance compared to the LPF--CSD1 and LPF--CSD2 methods with median correlation of $0.28$ ($95\%$ CI: $0.26$ to $0.30$)- Fig. 2A. Mean square error of LPF--CSD1 and LPF--CSD2 did not differ significantly; median error for LPF--CSD1 and LPF--CSD2 were $0.97$ ($95\%$ CI: $0.86$ to $1.07$) and $ 0.84$ ($95\%$ CI: $0.75$ to $0.93$) respectively, $P > 0.1$- Fig. 2B. 

The post-processing component improved the performance of the SSA-–DCT method such that SSA--DCT2 outperformed SSA--DCT1. The SSA--DCT2 method had a significantly higher correlation coefficient compared to SSA--DCT1 method: median correlation $0.86$ ($95\%$ CI: $0.83$ to $0.88$) in SSA--DCT2 compared to median correlation $0.6$ ($95\%$ CI: $0.57$ to $0.63$) in SSA--DCT1 method, $P < 0.001$. In addition, the SSA--DCT1 method had significantly higher median error compared to SSA--DCT2, $5.8$ (CI of $5.55$ to $6.06$) vs $1.48$ (CI of $1.33$ to $1.63$), $P < 0.001$ respectively- Fig. 2.
The comparison of SSA--DCT2 and LPF--CSD1 revealed that LPF--CSD1 outperformed SSA--DCT2 with a higher correlation coefficient and lower mean-squared error ($P < 0.001$). An example of these two methods for a synthetic signal is shown in Fig. 3. Both methods appear to remove all transients (Fig. 3A); however, on closer inspection (Fig. 3B and 3C), we find that neither method is perfect. LPF--CSD1 tends towards a very sparse solution, and undershoots the transients; in contrast, SSA--DCT2 does a better job in estimating the amplitude of the transients, but also tends to overshoot around the edge of the transients, possibly due to the filtering operation of the SSA--DCT2.

\section{DISCUSSION AND CONCLUSION}
 
This study sought to identify the best method to isolate and extract transient deflections within synthetic NIRS signals. We report that LPF--CSD1 was an effective and efficient method to isolate predefined transients in synthetic NIRS signals. Not only does the LPF--CSD1 perform better than LPF--SSA2 on synthetic NIRS signals, it also is computational efficient in comparison to the SSA-–DCT method, as is approximately 3 to 4 orders of magnitude faster than the SSA-–DCT methods. We found that the LPF–-CSD1 method outperforms other methods for decomposing the transient components in synthetic preterm NIRS-like signals. Although the post-processing component of the SSA–-DCT considerably improved performance of this method (correlation of $0.86$ compared to $0.6$), it did not outperform the best LPF-–TVD method. Performance of LPF-–CSD1 appears robust to a range of lambda values (see Fig. 2), nevertheless the optimization of these parameters may have provided this method with an advantage  over  SSA-–DCT2. Future studies could consider optimizing parameters of the SSA–-DCT method, although the computational overhead for this method may hinder a brute-force grid search. LPF–-CSD2 had been proposed as an improved version of LPF–-CSD1, in particular with estimating transient components, however we did not find this to be the case. Although we used a procedure to estimate these 4 parameters through a SD estimate of the noise as previously reported \cite{c6} and optimized one parameter, this method may not be optimal for synthetic NIRS signals that have a level of convexity. Furthermore, the non-convex penalty functions used in LPF-–CSD2 can have local minima which makes it challenging to find the global minimum \cite{c6}. However, we have identified a limitation of the LPF--CSD1 method which is not evident in the SSA--DCT2 method. Preliminary analysis suggests that SSA--DCT2 identifies transient amplitude more effectively than the LPF--CSD1 method. The amplitude of the NIRS signal is an important component of NIRS-guided therapy in the on-going clinical trial \cite{c2} therefore, extracting the full amplitude of transient oxygen desaturations is extremely important. Future studies will explore the suitability of LPF-—CSD1 and SSA—-DCT2 methods on larger datasets of real NIRS signals and investigate the role of cerebral oxygenation desaturations in preterm brain injury.

\addtolength{\textheight}{-12cm}   




\end{document}